\documentclass[12pt,twoside,A4,reqno]{amsart}
\usepackage{epsfig}

\calclayout

\def\({\left(}
       \def\){\right)}
       \def\[{\left[}
       \def\]{\right]}
    \def\<{\langle}
    \def\>{\rangle}

\newcommand{\bea}{\begin{eqnarray}}
\newcommand{\eea}{\end{eqnarray}}
\begin{document}
\title
  {Linearized feedforward control\\ of two-level quantum system
 by\\ modulated external field}
\maketitle
\begin{center}\author{SERGEI BORISENOK$^{1,\,\, 2}$,\,\, SAIFULLAH$^1$}\\
\vspace{.7cm} {$^1$ Abdus Salam School of Mathematical Sciences\\
Government College University\\
35 C -II, Gulberg III, Lahore, Pakistan\\
E-mail: saifullahkhalid75@yahoo.com
 \vspace{0.5cm}

$^2$ Dept. of Physics, Herzen State Pedagogical University\\
48 Moika River, 191186 St. Petersburg, Russia}
\end{center}

\begin{abstract}
We propose a model of feedforward (open-loop) optical control of
two-level atom in the linearized form. This model allows to express
the general form of solution for the atomic level populations via
the arbitrary shapes of the control signal. Then we make numerical
investigations of different shapes for the optical control signal.
\\

\noindent{\it PACS:} 02.30.Yy, 42.50.p \\

\noindent  {\it Key words: Feedforward control; Quantum optics}
\end{abstract}
\maketitle

\pagestyle{myheadings} \markboth{\centerline {\scriptsize SERGEI
BORISENOK,\,\, SAIFULLAH}} {\centerline {\scriptsize Linearized
Feedforward Control of Two-Level Quantum System by External Field}}
\section{Introduction}

\noindent A wide spectrum of control methods can be discovered for
the quantum systems. Among them feedforward (open-loop) approach
seems to be the most natural, since an applied external field, can
be easily designed as a time-dependent function. Here we will
discuss the basic, but very important case of two-level atomic
system controlled by modulated optical field. Our choice has been
motivated by developed technique for à practical design of external
field in quantum  optics.\\\\
\noindent Recently other authors studied the control of two-level
atoms in the frame of open loop-ideology when the controlling field
was known {\it a priori}. It allowed to get the different forms of
atomic energy spectra, producing $\pi$- and $\pi$/2-pulses
\cite{Imoto}, including the observation of the geometric phase using
stimulated photon echoes \cite{Tian}, taking special non-constant
shapes of external field  \cite{Di Piazza} etc.\\\\
\noindent We propose a model of feedforward control for the density
matrix in the linearized form. We use the \lq\lq semiclassical
approach\rq\rq\, of the atom--field interaction, when a single
quantum two-level atomic system (all other levels are neglected) is
interacting with classical electromagnetic field. We use the
standard notation following \cite{Scully}, but in our model the
optical field plays the role of a control signal $u(t)$ for
open-loop (feedforward) control scheme \cite{Fradkov}. A similar
case for the probability amplitudes (without decay) is described in
the frame of closed-loop scheme in \cite{Saifullah}. The present
model has a decay component, because it involves the effect of
elastic collisions between atoms.\\\\
\noindent In Section 2 we present our dynamical model with atomic
level population decay in generalized dimensionless form and then
apply the linearized control procedure for different shapes of the
optical control field $u(t)$. This model allows to express the
general form of solution for the atomic level populations via the
arbitrary shapes of the control signal. Then in Section 3 we make
numerical investigations of different shapes for the signal $u$.

\section{Feedforward optical control for two-level atom}

\subsection{Dynamical control model for two-level atom in classical\\
optical field:}

We consider the quantum two-level atomic system in the classical
optical field $E(t)$ (see Figure.1 in Appendix). Let $| a\rangle$
and $|b\rangle$ represent the upper and lower level states of the
atom, i.e., they are eigenstates of the unperturbed part of the
Hamiltonian $\hat H_0$ with the eigenvalues: $\hat
H_{0}|a\rangle=\hbar\omega_a|a\rangle$ and $\hat
H_{0}|b\rangle=\hbar\omega_b|b\rangle$.\\\\
\noindent The equations of motion for the density matrix elements
are given by \cite{Scully}:
\begin{eqnarray}
\nonumber \dot \rho _{aa}&=&-\gamma _a\rho _{aa}
+\frac{iE}{\hbar}\left(
\wp _{ab}\rho_{ba}e^{i\omega t}-\wp _{ab}^*\rho_{ab}e^{-i\omega t}\right) \ \ ;\\
\label{1} \dot \rho _{bb}&=&-\gamma _b\rho _{bb}
-\frac{iE}{\hbar}\left(
\wp_{ab}\rho_{ba}e^{i\omega t}-\wp _{ab}^*\rho _{ab}e^{-i\omega t}\right) \ \ ;\\
\nonumber \dot \rho _{ab}&=&-\gamma
_{ab}\rho_{ab}-\frac{iE}{\hbar}\wp _{ab} (\rho _{aa}-\rho
_{bb})e^{i\omega t} \ ,
\end{eqnarray}
where $\rho _{ba}=\rho _{ab}^*$; $\wp_{ab}$ is the matrix element of
the electric dipole moment, $\gamma _a$ and $\gamma _b$ are the
decay constants, $\gamma _{ab}=(\gamma _a +\gamma _b)/2+\gamma
_{ph}$ , $\gamma _{ph}$ is a decay rate including elastic collisions
between atoms, and  $\omega = \omega _a - \omega _b$ is the atomic
transition frequency.\\\\
\noindent Let's denote $\wp_{ab}=|\wp_{ab}|e^{i\phi}$ and
\begin{eqnarray}
\nonumber
\rho _{+}&\equiv&\rho_{ba}e^{i(\omega t+\phi )}+\rho _{ab}e^{-i(\omega t+\phi )}\ ;\\
\label{2} \rho _{-}&\equiv&i\left[ \rho_{ba}e^{i(\omega t+\phi
)}-\rho _{ab}e^{-i(\omega t+\phi )} \right] \ .
\end{eqnarray}
Using (\ref{2}) we can re-write the system (\ref{1}) in the real
form:
\begin{eqnarray}
\nonumber \dot \rho _{aa} &=& -\gamma _{a}\rho _{aa}+\frac{|\wp
_{ab}|E}{\hbar}\cdot
\rho_{-} \ ;\\
\nonumber \dot \rho _{bb} &=& -\gamma _{b}\rho _{bb}-\frac{|\wp
_{ab}|E}{\hbar}
\cdot \rho_{-} \ ;\\
\label{3}
\dot \rho _{+}&=&-\gamma _{ab}\rho _{+}+\omega \rho _{-}\ ;\\
\nonumber \dot \rho _{-}&=&-\gamma _{ab}\rho _{-}-\omega \rho
_{+}-\frac{2|\wp _{ab}|E}{\hbar} \cdot (\rho _{aa}-\rho _{bb}) \ .
\end{eqnarray}

\noindent For further calculations we put $\gamma _a = \gamma _b
\equiv \gamma$. Then
\begin{eqnarray}
\label{4} (\rho _{aa}+\rho _{bb})(t)=e^{-\gamma t}(\rho _{aa}+\rho
_{bb})(0) \ .
\end{eqnarray}
The first two equations of the system (\ref{3}) can be combined
together.\\ We can put:
\begin{eqnarray}
\nonumber
\rho _{aa}(t)-\rho _{bb}(t)&\equiv &e^{-\gamma t}x(t) \ ;\\
\label{5}
\rho _{+}(t)&\equiv &e^{-\gamma t}y(t) \ ;\\
\nonumber \rho _{-}(t)&\equiv &e^{-\gamma t}z(t) \ .
\end{eqnarray}
By substitution of (\ref{5}) in (\ref{3}) we can eliminate the decay
$\gamma$-containing terms. Finally, rescaling the time by $\omega$:
$\tau = \omega t$, and denoting the dimensionless control signal by
$u(t)\equiv 2|\wp _{ab}| E(t)/\hbar\omega$ and $\epsilon = \gamma
_{ph}/\omega$, we get the simplified system
\begin{eqnarray}
\nonumber
\dot x &=&u\cdot z\ ;\\
\label{6}
\dot y &=&-\epsilon \cdot y+ z\ ;\\
\nonumber \dot z &=& -\epsilon \cdot z -y-u\cdot x \ .
\end{eqnarray}
Here the dot means the derivative with respect to the new
dimensionless time $\tau$. We remind that $x \in [-1,1]$, since
$(\rho _{aa}-\rho_{bb} )\in [-1,1]$, and $(\rho _{aa}-\rho_{bb}) \to
0$ as $t \to \infty$.

\subsection{Linearization of control:}

Let's suppose that we apply the linearized form of control:
\begin{eqnarray}
\nonumber
x(\tau ) &=& X_0(\tau )+u\cdot X_1(\tau) \ ;\\
\label{7}
y(\tau ) &=& Y_0(\tau )+u\cdot Y_1(\tau) \ ;\\
\nonumber z(\tau ) &=& Z_0(\tau )+u\cdot Z_1(\tau) \ .
\end{eqnarray}

\noindent We will skip all the terms of the order $u^2$ and elder.
Then substituting (\ref{7}) in (\ref{6}), we split our system into
two parts: the free (non-controlled) system:
\begin{eqnarray}
\nonumber
\dot X_0 &=&0 \ ;\\
\label{8}
\dot Y_0 &=&-\epsilon \cdot Y_0+ Z_0\ ;\\
\nonumber \dot Z_0 &=& -\epsilon \cdot Z_0 -Y_0
\end{eqnarray}
and the controlled part:
\begin{eqnarray}
\nonumber
\dot u \cdot X_1+u\cdot \dot X_1 &=&u\cdot Z_0\ ;\\
\label{9}
\dot u \cdot Y_1+u\cdot \dot Y_1 &=&u \cdot Z_1\ ;\\
\nonumber \dot u \cdot Z_1+u\cdot \dot Z_1 &=& -u \cdot Y_1 -u\cdot
X_0 \ .
\end{eqnarray}

\noindent In (\ref{9}) we omitted the decay $\epsilon$-terms,
because the decay is supposed to be a slow process to compare with
the control, i.e. $\epsilon$ and $u$ are the small parameters of the
same order, and the linearization deals only with their first
orders. Then from the first equation of system (\ref{9}) we get:
$$
u(\tau )X_1(\tau )=\int_{0}^{\tau }u(t')Z_0(t')dt'
$$
and from the first equation of system (\ref{7}), we have
\begin{equation}
\label{10} x(\tau) = X_0(\tau) + \int_{0}^{\tau }u(t')Z_0(t')dt' \ .
\end{equation}
Now we apply the initial conditions $X_0(0)$, $Y_0(0)$, $Z_0(0)$ to
solve the system (\ref{8}):
\begin{eqnarray}
\nonumber
X_0(\tau ) &=& X_0(0) \equiv x(0)\ ;\\
\label{11} Y_0(\tau ) &=&e^{-\epsilon \tau} \left[
Y_0(0)\cos \tau + Z_0(0)\sin \tau \right] \ ;\\
\nonumber Z_0(\tau ) &=&e^{-\epsilon \tau} \left[ Z_0(0)\cos \tau -
Y_0(0)\sin \tau \right] \ .
\end{eqnarray}

\noindent If we denote the phase of $\rho _{ab}$ by $\phi '$, then
$\rho _{+}(0)=2|\rho _{ab}|\cos (\phi '- \phi )$ and $\rho
_{-}(0)=2|\rho _{ab}|\sin (\phi '- \phi )$. We can put for the
initial condition: $\phi ' = \phi$, then $\rho _{+}(0)=2|\rho
_{ab}(0)|\equiv \delta$ and $\rho _{-}(0)=0$. Let's demand
$X_1(0)=Y_1(0)=Z_1(0)=0$. Thus, $Y_0(0)=\delta$ and $Z_0(0)=0$ are
our initial conditions.

\subsection{Control signal correction:}

If $X_0(0)=-1$ (that corresponds to the ground level of the atom as
the initial condition), then from $-1\leq x(\tau) \leq 1$ and
(\ref{10}) it follows:
\begin{equation}
\label{12} 0\leq \int_{0}^{\tau }u(t')Z_0(t')dt' \leq 2.
\end{equation}
In other words this integral should be positive and bounded. We
define first the arbitrary non-corrected control $u_0(\tau )$ and
then put
\begin{equation}
\label{13} \tilde u(\tau )\equiv |u_0(\tau )|\cdot
\textrm{sign}Z_0(\tau )\ .
\end{equation}
Then the left inequality (\ref{12}) will be satisfied automatically.
The right part of (\ref{12}) can be represented by Cauchy --
Schwartz inequality:
$$
\left| \int_{0}^{\tau }\tilde u(t')Z_0(t')dt' \right| ^2 \leq
\int_{0}^{\tau }\tilde u^2(t')dt' \cdot \int_{0}^{\tau
}Z_0^2(t'')dt'' \ ,
$$
and then we demand:
\begin{equation}
\label{14} \int_{0}^{\tau }\tilde u^2(t')dt' \cdot \int_{0}^{\tau
}Z_0^2(t'')dt'' \leq 4 \ .
\end{equation}
Let's check the inequality (\ref{14}):
\begin{eqnarray}
\nonumber \left| \int_{0}^{\tau}Z_0^2(t'')dt'' \right| = |Y_0(0)|^2
\left| \int_{0}^{\tau}e^{-2\epsilon t''}\sin ^2 t'' dt'' \right| \leq \\
\label{15} \leq \delta ^2 \left| \int_{0}^{\tau}e^{-2\epsilon t''}
dt'' \right|= \frac{\delta ^2 (1-e^{-2\epsilon \tau})}{2\epsilon} \
.
\end{eqnarray}
Thus, from (\ref{14}) and (\ref{15})
\begin{equation}
\label{16} \int_{0}^{\tau} \tilde u^2(t') dt' \leq \frac{8\epsilon
}{\delta ^2(1-e^{-2\epsilon \tau})} \ .
\end{equation}

\noindent To satisfy (\ref{16}) we also have to correct the signal
$u$. Let's suppose that there are two functions: an initial
arbitrary $u_0(\tau )$ and its corrected variant $u(\tau )$ that is
bounded above by the condition (\ref{16}). Of course, physically the
external optical field should follow the signal $u$, and the initial
$u_0$ is only a basic model to construct the behavior of the
open-loop control field.\\\\
\noindent Now let's define
\begin{equation}
\label{17} \Delta (\tau ) \equiv \int_{0}^{\tau} \tilde u^2(t') dt'
- \frac{8\epsilon }{\delta ^2(1-e^{-2\epsilon \tau})}
\end{equation}
and
\begin{eqnarray}
\label{18}
 u (\tau )&=& \left\{
  \begin{array}{ll}
    \tilde u(\tau ) & \hbox{$\ , \ \Delta(\tau )< 0\ ;$} \\
    B(\tau ) & \hbox{$\ ,\ \Delta (\tau )\geq 0 \ ,$}
      \end{array}
\right.
\end{eqnarray}
where a positive function $B(\tau )$ is defined from the next
equation:
\begin{equation}
\label{19} \int_{0}^{\tau} B^2(t') dt' = \frac{8\epsilon }{\delta
^2(1-e^{-2\epsilon \tau})} \ ,
\end{equation}
or
$$
B^2(\tau ) = \left| \frac{d}{d\tau}\ \frac{8\epsilon }{\delta
^2(1-e^{-2\epsilon \tau })} \right| =\frac{16\epsilon ^2}{\delta
^2}\cdot\frac{e^{-2\epsilon \tau}}{(1-e^{-2\epsilon \tau})^2}\ .
$$
Thus,
\begin{equation}
\label{20} B(\tau) = \frac{4\epsilon}{\delta}\cdot\frac{e^{-\epsilon
\tau}}{1-e^{-2\epsilon \tau }}=\frac{2\epsilon}{\delta \cdot
\textrm{cosh}(\epsilon \tau)}\ .
\end{equation}
For small time intervals $\tau << 1/\epsilon$ we have: $\delta ^2
\tau$ in RHS (\ref{15}), and\\ $B(\tau) \simeq 2/(\delta \cdot
\tau)$.\\\\
\noindent Finally by the corrections (\ref{13}) and (\ref{18}) we
have:
\begin{eqnarray}
\label{21} x(\tau) = -1 + \delta \cdot \int_{0}^{\tau }dt'\
e^{-\epsilon t'}|\sin t'| \cdot |u(t')| \ .
\end{eqnarray}
Eq.(\ref{21}) solves the problem of open-loop control in linearized
form. Now defining the control signal $u(\tau )$ we restore by
(\ref{21}) the shape of the difference $\rho _{aa}(t)-\rho
_{bb}(t)$. Their sum (\ref{4}) is known, thus, we can find $\rho
_{aa}(t)$ and $\rho _{bb}(t)$ separately.

\section{Numerical simulation of different shapes for the control
signal} \noindent Now we can apply the general solution of
Eq.(\ref{21}) to study the influence of control optical field $u$ on
the behavior
of the system (\ref{6}).\\\\
\noindent In the case of an ideal open-loop control the behavior of
$x(t)$ is the saturation of the population at the ground level, in
other words, $x(t) \to 1$ as $t \to \infty$. Sure, not every control
will satisfy this condition.\\\\
\noindent On. Figs. 2--6 (see Appendix) we plot the different shapes
of the initial $u_0$ (Figs.A) and corrected $u$ (Figs.B) control
signals: constant, ramp, step, sine wave, and repeating sequence
stair. To compare their efficiency we check also the corresponding
time derivatives $dx/dt$ (Figs.D).\\\\
\noindent We can see from the plots that the ramp control in our
case is definitely more effective. The speed of the saturation for
$x(t)$ is faster for the signals on Figs. 3, 6.
\section{Conclusion}
\noindent Finally we can conclude that our model for the open-loop
control has
several important features: \\\\
1. It can be easily extended for the case of multi-level atomic
systems by adding the correspondent components in the density
matrix;\\\\
2. For the two-level system it can be re-formulated in general form
if we propose the linear approximation of control;\\\\
3. It can be an origin of studying the behavior of controlled
non-linear systems in quantum optics.
\section{Acknowledgement}
\noindent We are very grateful to Dr.Yuri Rozhdestvensky (Institute
of Laser Physics, Saint Petersburg, Russia.) for productive
scientific discussions.

\newpage
\section{Appendix: Figures}

\begin{figure}[htbp]
 \centerline{\includegraphics[width=8.0cm]{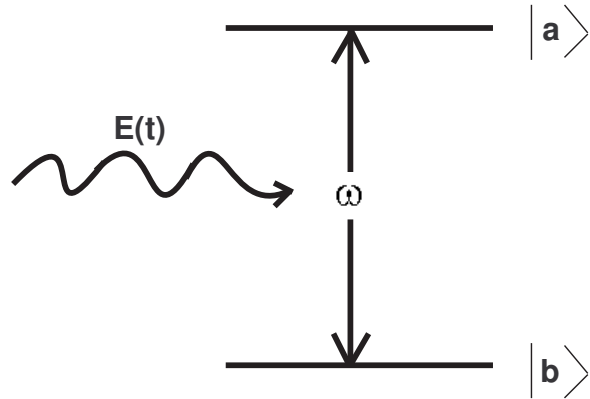}}
   \label{fig Ex1}
    \caption{Interaction of a single two-level atom with an optical field.}
\end{figure}

\begin{figure}[htbp]
 \centerline{\includegraphics[width=10.7cm]{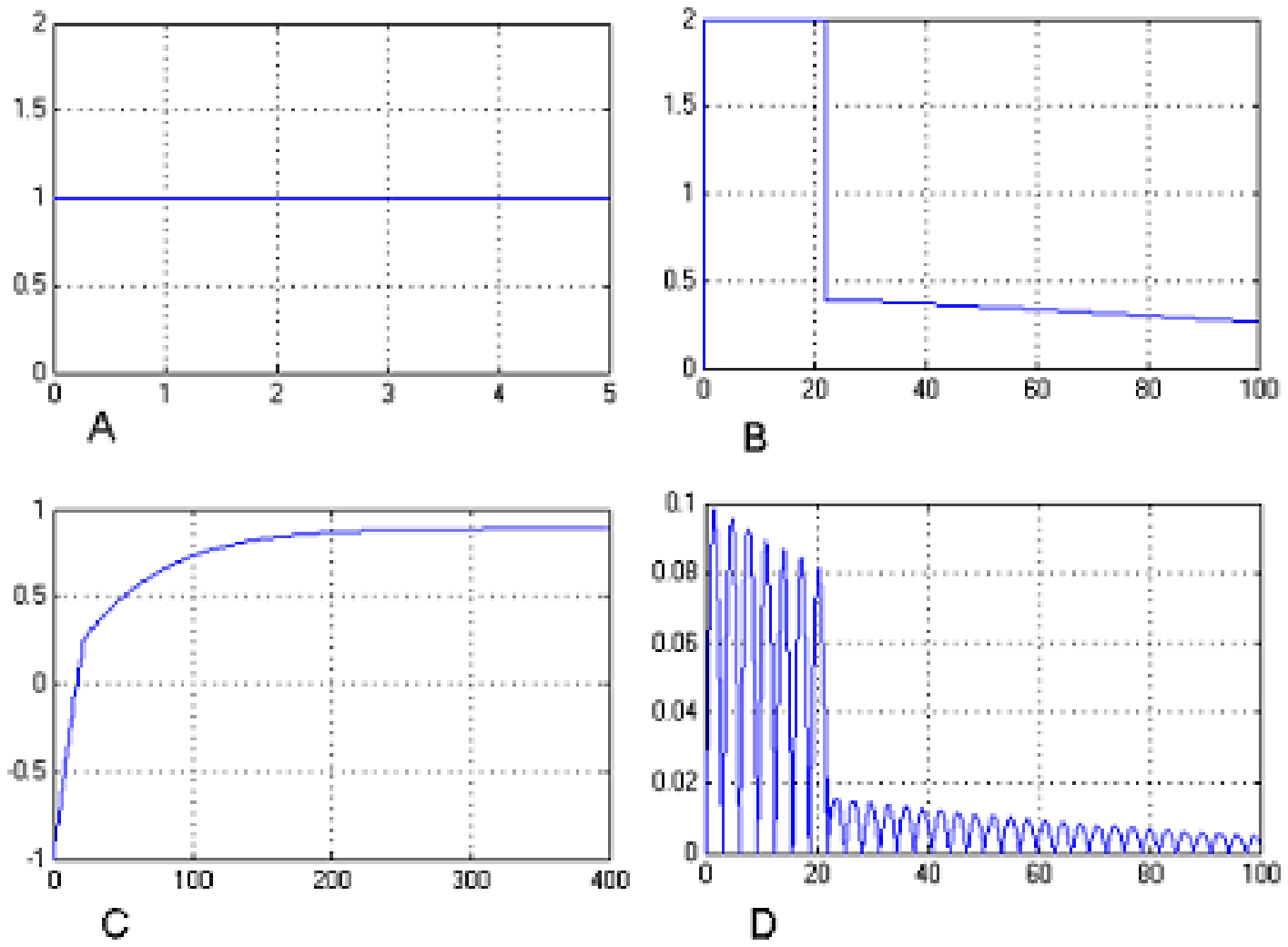}}
   \label{fig Ex1}
    \caption{Constant control signal: (A) The initial signal $u_0(t)$;
    (B) The corrected signal $u(t)$; (C) $x(t)$; (D) The derivative $dx/dt$.}
\end{figure}

\begin{figure}[htbp]
 \centerline{\includegraphics[width=10.7cm]{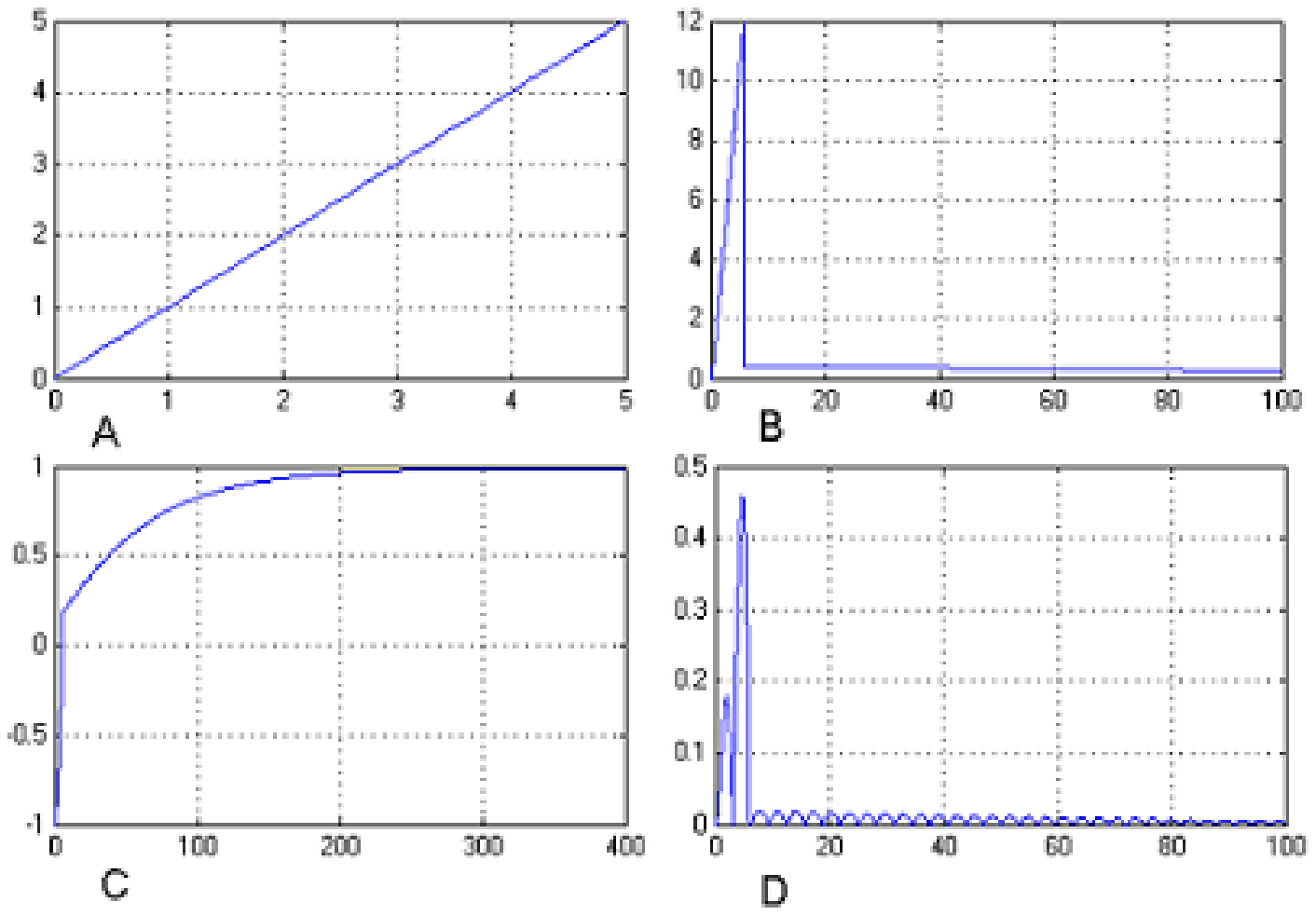}}
   \label{fig Ex1}
    \caption{Ramp control signal: (A) The initial signal $u_0(t)$;
    (B) The corrected signal $u(t)$; (C) $x(t)$; (D) The derivative $dx/dt$.}
\end{figure}

\begin{figure}[htbp]
 \centerline{\includegraphics[width=10.7cm]{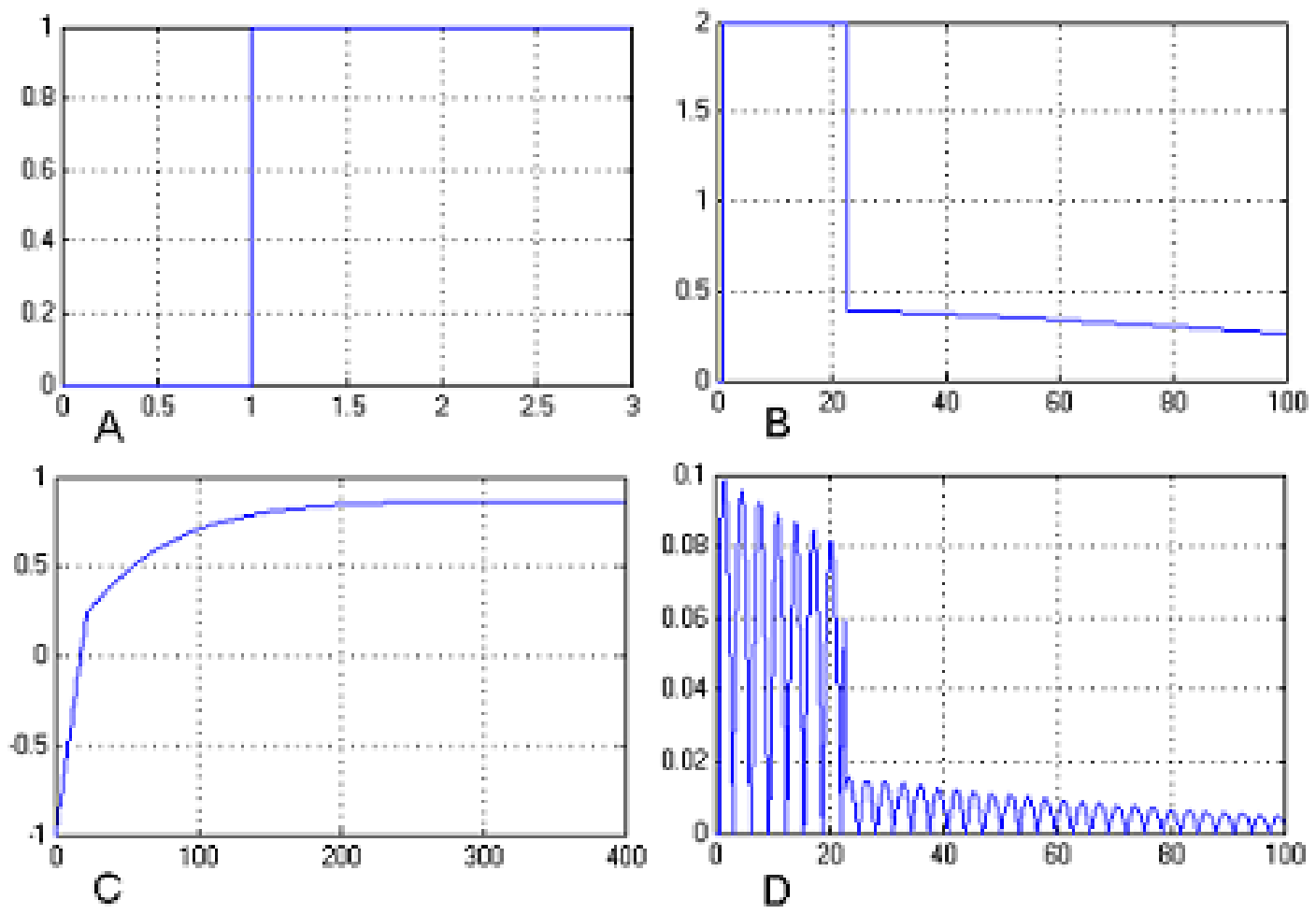}}
   \label{fig Ex1}
    \caption{Step control signal: (A) The initial signal $u_0(t)$;
    (B) The corrected signal $u(t)$; (C) $x(t)$; (D) The derivative $dx/dt$.}
\end{figure}

\begin{figure}[htbp]
 \centerline{\includegraphics[width=10.7cm]{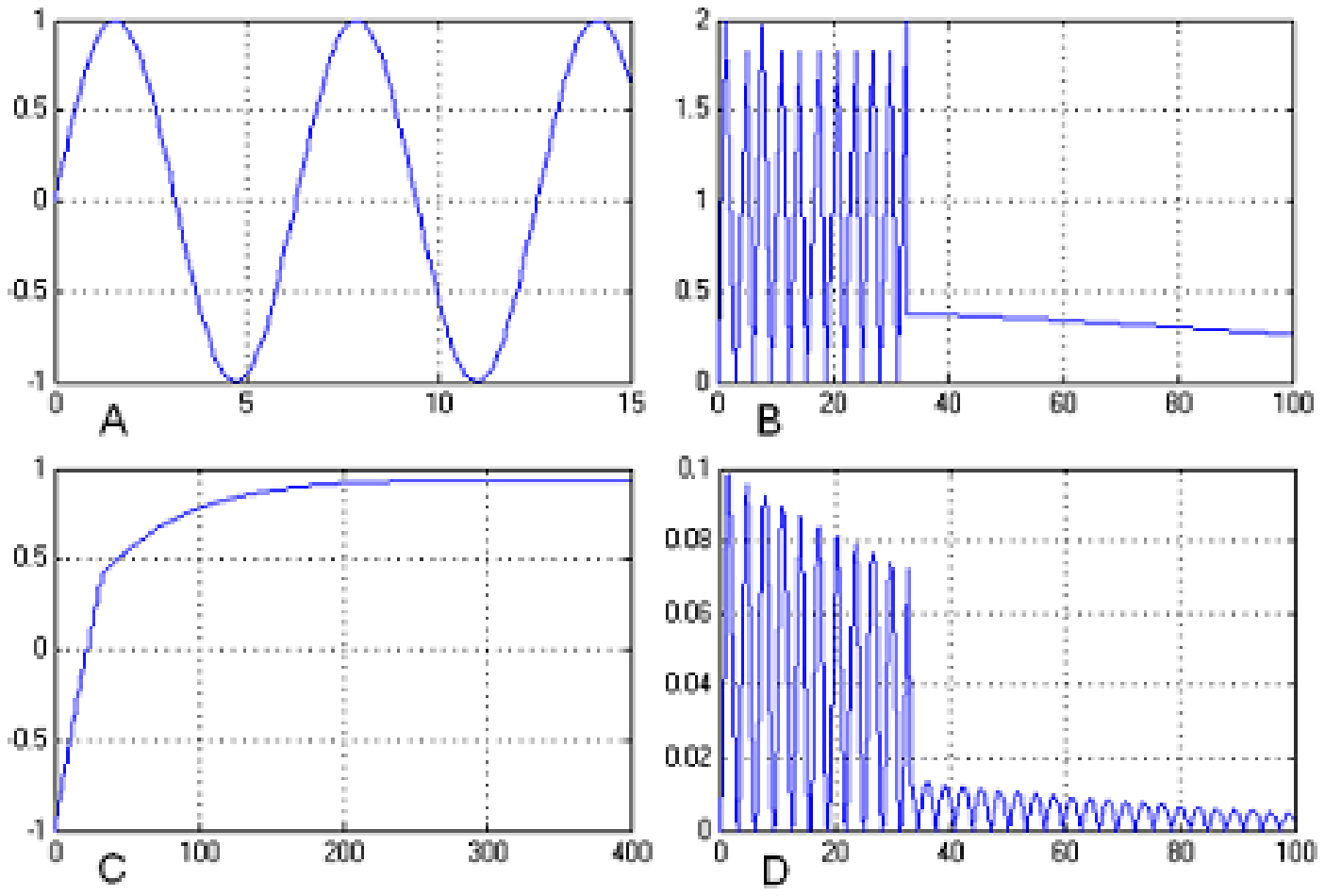}}
   \label{fig Ex1}
    \caption{Sine wave control signal: (A) The initial signal $u_0(t)$;
    (B) The corrected signal $u(t)$; (C) $x(t)$; (D) The derivative $dx/dt$.}
\end{figure}

\begin{figure}[htbp]
 \centerline{\includegraphics[width=10.7cm]{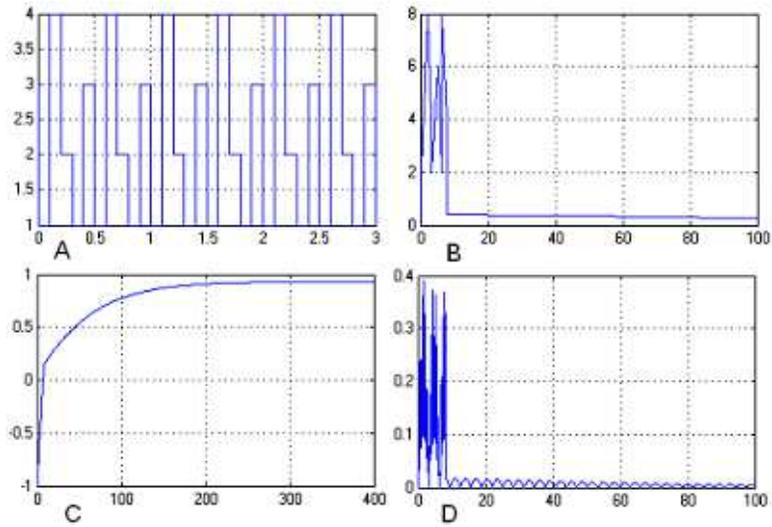}}
   \label{fig Ex1}
    \caption{Repeating sequence stair control signal: (A) The initial signal $u_0(t)$;
    (B) The corrected signal $u(t)$; (C) $x(t)$; (D) The derivative $dx/dt$.}
\end{figure}
\end{document}